# Frequency Quantized Nondiffracting X Waves in a Confined Space

Jian-yu Lu, Ph.D.

Ultrasound Laboratory, Department of Bioengineering, The University of Toledo, Toledo, OH 43606, U.S.A. E-mail: jilu@eng.utoledo.edu, Web: http://www.eng.utoledo.edu/~jilu (or search by author' name on web)

*Abstract* - This paper presents theoretical results indicating that newly discovered nondiffracting beams we call X waves, can propagate in a confined space (wave guide) with specific quantized temporal frequencies. These results could have applications in nondispersive transmission of acoustic, electromagnetic (microwaves) and optical energy through wave guides as well as provide new localized wave functions of de Broglie waves in quantum mechanics.

## I. INTRODUCTION

Recently, families of generalized solutions to the isotropic/homogeneous scalar wave equation which represent nondiffracting waves have been discovered (Lu and Greenleaf, 1992a). These nondiffracting waves are not physically realizable because infinite aperture and energy are required to produce them. However, approximate nondiffracting waves can be produced over deep depth of field by truncating them in both space and time resulting in possible applications in medical imaging and tissue characterization (Lu and Greenleaf, 1990a, 1990b, 1991, 1992b). In this paper, it is shown that temporal-frequency quantized nondiffracting X waves can propagate in a confined space such as a wave guide, normally highly dispersive devices under appropriate boundary conditions. This would allow nondispersive transmission of energy. Shaarawi et al. (1989) and Ziolkowski et al. (1991) have shown that the "localized waves" such as Focused Wave Modes and Modified Power Spectrum Pulses, etc., can also propagate in wave guides for an extended propagation depth. These results are important because they could have applications in acoustic and electromagnetic energy transmission, seismology, underwater acoustics, fiber optics and microwaves, etc.

## II. ACOUSTIC WAVES

The propagation of acoustic waves in three-dimensional isotropic/homogeneous media is governed by the following scalar wave equation

$$\left[\frac{1}{r}\frac{\partial}{\partial r}(r\frac{\partial}{\partial r}) + \frac{1}{r^2}\frac{\partial^2}{\partial \phi^2} + \frac{\partial^2}{\partial z^2} - \frac{1}{c^2}\frac{\partial^2}{\partial t^2}\right]\Phi = 0, \qquad (1)$$

where $r = \sqrt{x^2 + y^2}$ represents radial coordinate, $\phi$ is azimuthal angle, $z$ is axial axis, which is perpendicular to the plane defined by $r$ and $\phi$, $t$ is time, $c$ is the speed of sound of the media and $\Phi$ represents acoustic pressure or velocity potential which is a function of $r$, $\phi$, $z$ and $t$.

One of the families of the generalized solutions to (1) is given by (Lu and Greenleaf, 1992a):

$$\Phi_\zeta(s) = \int_0^\infty T(k)\left[\frac{1}{2\pi}\int_{-\pi}^{\pi} A(\theta)f(s)d\theta\right]dk, \qquad (2)$$

where

$$s = \alpha_0(k,\zeta)r\cos(\phi-\theta) + b(k,\zeta)\left[z \pm c_1(k,\zeta)t\right], \qquad (3)$$

and where

$$c_1(k,\zeta) = c\sqrt{1+\left[\alpha_0(k,\zeta)/b(k,\zeta)\right]^2}, \qquad (4)$$

where $T(k)$ is any complex function (well behaved) of $k$ and could include the temporal frequency transfer function of a radiator system, $A(\theta)$ is any complex function (well behaved) of $\theta$ and represents a weighting function on $\theta$, $f(s)$ is any complex function (well behaved) of $s$, and $\alpha_0(k,\zeta)$ and $b(k,\zeta)$ are any complex function of $k$ and $\zeta$.

If $c_1(k,\zeta)$ in (4) is real, "$\pm$" in (3) represent the waves propagating along the negative and positive $z$ directions, respectively (without loss of generality, in the following we consider only waves along positive $z$ direction). In addition, $\Phi_\zeta(s)$ in (2) represents a family of nondiffracting waves if $c_1(k,\zeta)$ is independent of the parameter, $k$.

Selecting the following specific functions, $T(k) = B(k)e^{-a_0 k}$, $A(\theta) = i^n e^{in\theta}$, $f(s) = e^s$, $\alpha_0(k,\zeta) = -ik\sin(\zeta)$, and $b(k,\zeta) = ik\cos(\zeta)$, we obtain the nth-order nondiffracting X waves (Lu and Greenleaf, 1992a):

$$\Phi_{X_n} = e^{in\phi}\int_0^\infty B(k)J_n(kr\sin\zeta)e^{-k[a_0 - i\cos\zeta(z-c_1 t)]}dk, \ (n=0, 1, 2, ...), \tag{5}$$

where $B(k)$ is any complex function (well behaved) of $k$ and represents the temporal frequency transfer function of a radiator system, $k=\omega/c$ and $\zeta \leq \pi/2$ are parameters which are independent of the spatial position, $\vec{r} = (r\cos(\phi), r\sin(\phi), z)$, and time, $t$, $\omega$ is angular frequency, $J_n$ is the nth-order Bessel function of the first kind, $a_0 > 0$ is a constant and $c_1 > c/\cos\zeta$ is the propagation speed of the X waves.

If $\Phi$ in (1) representing acoustic pressure, an infinitely long cylindrical acoustical wave guide of a diameter of $a$ with an infinite rigid boundary which is filled with an isotropic/homogeneous lossless fluid medium can be used as an example. In this case, the normal vibration velocity of the medium at the wall of the cylindrical wave guide will be zero for all the frequency components of the X waves, i.e., $\dfrac{i}{\rho_0 \omega}\dfrac{\partial}{\partial r}\Phi_{X_n}(\vec{r},t;\omega) \equiv 0$, $\forall \omega \geq 0$ at $r=0$, where $\rho_0$ is the density of the medium at equilibrium state and $\Phi_{X_n}(\vec{r},t;\omega)$ is the X wave component at the angular frequency, $\omega$ (see (5)). To meet this boundary condition, the parameter, $k$, in (5) should be quantized:

$$k_{nj} = \frac{\mu_{nj}}{a\sin\zeta}, \ (n,j=0, 1, 2, ...), \tag{6}$$

where $\mu_{nj}$ are the roots of the equations

$$\begin{cases} J_1(x) = 0, & n = 0 \\ J_{n-1}(x) = J_{n+1}(x), & n = 1, 2, ... \end{cases} \tag{7}$$

For the wave guide, the integral in (5) can be changed to a series representing frequency quantized nondiffracting X waves:

$$\Phi_{X_n}(\vec{r},t) = e^{in\phi} \sum_{j=0}^{\infty} \Delta k_{nj} B(k_{nj}) J_n(k_{nj} r \sin \zeta) e^{-k_{nj}[a_0 - i\cos\zeta(z-c_1 t)]}, \ r \leq a, \tag{8}$$

$$(n = 0, 1, 2, ...)$$

where $\Delta k_{n0} = k_{n1}$ and $\Delta k_{nj} = k_{nj+1} - k_{nj}$ ($j = 1, 2, 3, ...$). Unlike conventional guided waves, the frequency quantized nondiffracting X waves contain multiple frequencies and propagate through wave guides at the speed of $c_1$ without dispersion. It is noticed that similar results can also be obtained for wave guides with other homogeneous boundary conditions such as an infinitely long cylindrical acoustical wave guide with a diameter of $a$ which is composed of isotropic/homogeneous lossless media in a free space (vacuum). In this case, the acoustical pressure is zero at the boundary of the wave guide, $r = a$, i.e., $\mu_{nj}$, ($n, j = 1, 2, 3, ...$), in (6) are roots of $J_n(x) = 0$, ($n = 1, 2, ...$). See Figs. 1 to 3 for an example of the waves in an acoustic waveguide.

Now we discuss briefly the implications of (8). If $n = 0$, (8) represents an axially symmetric frequency quantized nondiffracting X wave. Because $B(k_{nj})$ in (8) is arbitrary, a source with any system transfer function can be used to excite the frequency quantized nondiffracting X waves. The X waves in (8) have a finite transverse spatial extension because the radius, $a$, is finite. Because the frequency quantized nondiffracting X waves are a superposition of CW waves with various frequencies, they are noncausal.

The frequency quantized nondiffracting X waves will become nondiffracting X waves in the limit, that is, if $a \to \infty$, then $\Delta k_{nj} \to 0$ and the summation in (8) becomes an integration which represents the nondiffracting X waves. On the other hand, if the confined space is reduced, i.e., $a \to 0$, both $k_{nj}$ and $\Delta k_{nj} \to \infty$ $(n, j = 0, 1, 2, ...)$. This means that for a small wave guide, only high frequency quantized nondiffracting X waves can propagate through it.

Additional nondiffracting waves with quantized parameters could also propagate in the confined space. For instance, let $\Phi_\zeta(s|_{r=a}) \equiv 0$ or $\frac{\partial}{\partial r}\Phi_\zeta(s|_{r=a}) \equiv 0$, we have (see (2) and (3)):

$$\frac{1}{2\pi}\int_{-\pi}^{\pi} A(k)f(s|_{r=a})d\theta \equiv 0 \text{ or } \frac{1}{2\pi}\int_{-\pi}^{\pi} A(k)\frac{\partial}{\partial r}f(s|_{r=a})d\theta \equiv 0, \tag{9}$$

or

$$f(s|_{r=a}) \equiv 0 \text{ or } \frac{\partial}{\partial r}f(s|_{r=a}) \equiv 0. \tag{10}$$

If $c_1$ in (4) is real and independent of $k$, and (9) and (10) are satisfied for discrete $k$ and $\theta$, (2) represents a family of generalized parameter-quantized nondiffracting waves by replacing the integrals with series.

## II. ELECTROMAGNETIC WAVES

The free-space vector wave equations (obtained from the free-space Maxwell's equations, Meyer-Arendt, 1972) are given by:

$$\nabla^2 \vec{E} - \frac{1}{c^2}\frac{\partial^2 \vec{E}}{\partial t^2} = 0, \tag{11}$$

and

$$\nabla^2 \vec{H} - \frac{1}{c^2}\frac{\partial^2 \vec{H}}{\partial t^2} = 0. \tag{12}$$

where $\vec{E}$ and $\vec{H}$ are electric and magnetic field strength, respectively, and $\nabla^2$ is a Laplace operator. An solution to (11) can be written as:

$$\vec{E}(\vec{r},t) = \vec{E}(r,\phi)e^{\gamma z - i\omega t}, \tag{13}$$

where $\vec{r} = (x, y, z)$ represents a point in the space, $\omega$ is angular frequency, $\gamma = i\beta$ is a propagation constant, $\beta = \sqrt{k^2 - k_c^2} > 0$ (for propagation waves), $k = \omega/c$ is a wave number, $c$ is the speed of light, and $\vec{E}(r,\phi)$ is a solution of the transverse vector Helmholtz equation:

$$\nabla_\perp \vec{E}(r,\phi) + k_c^2 \vec{E}(r,\phi) = 0, \tag{14}$$

where $\nabla_\perp$ is the transverse Laplace operator and $k_c$ is a parameter which is independent of $r$, $\phi$, $z$, and $t$.

For transverse magnetic (TM) waves, from (14) we have:

$$\nabla_\perp E_z(r,\phi) + k_c^2 E_z(r,\phi) = 0, \tag{15}$$

where $E_z(r,\phi)$ is the $z$ component of the vector $\vec{E}(r,\phi)$.

If we let $k_c = k \sin \zeta$, where $0 < \zeta < \pi/2$ is a constant, we obtain solutions of (15) of order n;

$$E_{z_n}(r,\phi) = e^{in\phi} B(k) e^{-ka_0} J_n(kr \sin \zeta), \ (n = 0, 1, 2, ...). \tag{16}$$

Nth-order nondiffracting X wave solution can be constructed from Eq. (16) by including the exponential term of (13) and integrating over $k$ from 0 to $\infty$:

$$E_{z_{nX}}(\vec{r},t) = e^{in\phi} \int_0^\infty B(k) J_n(kr \sin \zeta) e^{-k[a_0 - i\cos\zeta(z - c_1 t)]} dk, \ (n = 0, 1, 2, ...), \tag{17}$$

where "X" means "X wave".

Because (17) is the same as (5), the frequency quantization procedures for (5) can also be applied to (17). Suppose that the electromagnetic X waves are traveled in vacuum in a totally conductive cylindrical wave guide with a radius of $a$ (i.e., $E_{z_{nX}}(\vec{r},t) \equiv 0$ at $r = a$), we have:

$$E_{z_{nX}}(\vec{r},t) = e^{in\phi}\sum_{j=0}^{\infty}\Delta k_{nj} B(k_{nj}) J_n(k_{nj} r \sin\zeta) e^{-k_{nj}[a_0 - i\cos\zeta(z - c_1 t)]}, \; r \leq a, \quad (18)$$

$$(n = 0, 1, 2, ...)$$

where $k_{nj}$ ($n, j = 0, 1, 2, ...$) are given by (6), and $\mu_{nj}$ ($n, j = 0, 1, 2, ...$) in (6) are roots of $J_n(x)=0$ ($n = 0, 1, 2, ...$). Other components of $\vec{E}$ and the magnetic field strength, $\vec{H}$, can be derived from $E_z(\vec{r},t)$ using the free-space Maxwell's equations. They have the same speed, $c_1$, as $E_z$. For transverse electric (TE) waves, results would be similar.

### III. DE BROGLIE'S WAVES

The general nonrelativistic, time-dependent, and three-dimensional Schrodinger wave equation is given by (Sandin, 1989):

$$-\frac{\hbar^2}{2m}\nabla^2\Phi + V\Phi = i\hbar\frac{\partial\Phi}{\partial t}, \quad (19)$$

where $\hbar = h/2\pi$ $h$ is the Plank constant, $m$ is mass of particle, $\Phi$ is wave function of the de Broglie's waves and $V$ is a potential energy function of the system. For free-particles, $V = 0$, we obtain:

$$-\frac{\hbar^2}{2m}\nabla^2\Phi = i\hbar\frac{\partial\Phi}{\partial t}. \quad (20)$$

It is easy to prove that if $f(s) = e^s$, (2) is an exact solution of (20), where

$$c_1(k,\zeta) = \frac{\hbar\left[a_0^2(k,\zeta) + b^2(k,\zeta)\right]}{i2mb(k,\zeta)}. \tag{21}$$

Let $a_0(k,\zeta) = -ik\sin\zeta$ and $b(k,\zeta) = ik\cos\zeta$, where we assume that $k = 2\pi/\lambda$ is wave number and $\lambda$ is the wavelength of the de Broglie's waves, from (21), we obtain:

$$c_1(k,\zeta) = \frac{\hbar k}{2m\cos\zeta}. \tag{22}$$

Because $\hbar k = p$, where $p = mv$ is momentum, and where $v$ is the speed of particle, (22) becomes

$$c_1(k,\zeta) = \frac{v(k)}{2\cos\zeta}. \tag{23}$$

Let $T(k) = B(k)e^{-a_0 k}$ and $A(\theta) = i^n e^{in\theta}$, and use (23), from (2), we obtain an nth-order nondiffracting X wave solution to (20)

$$\Phi^s_{X_n} = e^{in\phi}\int_0^\infty B(k) J_n(kr\sin\zeta) e^{-k[a_0 - i\cos\zeta(z - c_1 t)]} dk. \tag{24}$$

$(n = 0, 1, 2, ...)$

(24) is the same as (5) except that the expression for $c_1$ is different. This means that $\Phi^s_{X_n}$ could be a new wave function of a free particle in the free space.

With a finite transverse spatial extension (such as a free particle passing through a hole of finite aperture), the wave function $\Phi^s_{X_n}$ in (24) would change (spread or diffract) after certain distance behind the hole. However, if the nondiffracting X waves in (24) is quantized in temporal frequency under the boundary conditions discussed in the previous sections (such as particles pass through a pipe), the frequency quantized nondiffracting X wave functions of the particles would propagate in the confined space without modifying their complex waveforms:

$$\Phi^s_{X_n}(\vec{r},t) = e^{in\phi}\sum_{j=0}^{\infty}\Delta k_{nj}B(k_{nj})J_n(k_{nj}r\sin\zeta)e^{-k_{nj}[a_0-i\cos\zeta(z-c_1 t)]}, \ r \leq a, \quad (25)$$

$$(n = 0, 1, 2, ...)$$

Because the wavenumber, $k$, in (25) is quantized, the momentum, $p = \hbar k$; and the energy of the particles are quantized in the confined space. The probability of finding a particle in the confined space would be proportional to $\Phi^s_{X_n}(\vec{r},t)[\Phi^s_{X_n}(\vec{r},t)]^*$, where "*" represents complex conjugate.

## IV. DISCUSSION

*a. Possible Relationship between Waves and Particles*

Because of the fact that the frequency quantized nondiffracting X waves are localized and are also solutions of the nonrelativistic Shrodinger wave equation for free particles (the potential energy function of the system is zero), they might have some intrinsic relationship with particles. The free choice of the parameter, $a_0 > 0$, could increase the localization of the frequency quantized nondiffracting X waves.

*b. Interaction between Light and Materials*

The light in free space behaves like a wave, but acts as particles (photons) when interacts with materials. The microscopic structures of the materials could be considered as some kind of optical wave guides within which the light waves are confined. From the above discussion of the nondiffracting X waves in a confined space, it is understood that only the light waves which have higher energy (or frequency) can penetrate the materials to cause the interaction.

## V. SUMMARY

The generalized nondiffracting waves discovered recently can propagate in a confined space with quantization of some parameters. This is of interest because they are localized and could be applied to acoustic, electromagnetic, and optical wave guides for energy transmission.

NOTE

This paper was first written in July 25, 1991 by Dr. Jian-yu Lu who, at that time, was working at Mayo Clinic, Rochester, MN, USA. The paper was revised on March 7, 1992 after it was proof read by Dr. James F. Greenleaf. However, this paper has never been published. Upon the request by Dr. Erasmo Recami, this paper was sent to him on January 20, 2000. Later, he had acknowledged this paper in his 2001 Physical Review E paper (see the Acknowledgements Section below). To provide a reference, Dr. Jian-yu Lu is going to place this paper in arXiv unmodified from the version of March 7, 1992. Therefore, this paper represents the views and results (correct or incorrect) of Dr. Jian-yu Lu of that time.

ACKNOWLEDGEMENTS

The author would like to thank Dr. James F. Greenleaf, Mayo Clinic, Rochester, MN, USA for proof reading the manuscript. The author would also like to thank Dr. Erasmo Recami, Facolta' di Ingegneria, Universita' statale di Bergamo, Dalmine (BG), Italy; INFN, for his interest and acknowledgement of this paper in their paper: Michel Zamboni-Rached, Erasmo Recami, and Flavio Fontana, "Superluminal localized solutions to Maxwell equations propagating along a normal-sized waveguide," Physical Review E, v. 64, 066603 (2001), p. 5. This work was supported in part by grants CA 43920 and CA 54212–01 from the National Institutes of Health.

REFERENCES

Jian-yu Lu and J. F. Greenleaf, "Ultrasonic nondiffracting transducer for medical imaging," IEEE Transactions on Ultrasonics, Ferroelectrics, and Frequency Control, Vol. 37, No. 5, pp. 438–447, September, 1990a

Jian-yu Lu, and J. F. Greenleaf, "Evaluation of a nondiffracting transducer for tissue characterization," [IEEE 1990 Ultrasonics Symposium, Honululu, Hawaii, U.S.A., December

4–7, 1990], IEEE 1990 Ultrasonics Symposium Proceedings, 90CH2938–9, Vol. 2, pp. 795–798, 1990b

Jian-yu Lu and J. F. Greenleaf, "Pulse-echo imaging using a nondiffracting beam transducer," Ultrasound in Medicine and Biology, Vol. 17, No. 3, pp. 265–281, May, 1991

Jian-yu Lu and J. F. Greenleaf, "Nondiffracting X waves—exact solutions to free-space scalar wave equation and their finite aperture realizations," IEEE Transactions on Ultrasonics, Ferroelectrics, and Frequency Control, Vol. 39, No. 1, pp. 19–31, January 1992a

Jian-yu Lu and J. F. Greenleaf, "Experimental verification of nondiffracting X waves," IEEE Transactions on Ultrasonics, Ferroelectrics, and Frequency Control, Vol. 39, No. 3, May 1992b (to be published)

J. R. Meyer-Arendt, INTRODUCTION TO CLASSICAL AND MODERN OPTICS. Englewood Cliffs, NJ, Prentice-Hall, Inc., 1972, ch. 6

T. R. Sandin, ESSENTIALS OF MODERN PHYSICS. New York: Addison-Wesley Publishing Company, 1989, p.146

A. M. Shaarawi, I. M. Besieris, and R. W. Ziolkowski, "Localized energy pulse trains launched from an open, semi-infinite, circular wave guide," Journal of Applied Physics, Vol. 65, No. 2, pp. 805–813, January 15, 1989

R. W. Ziolkowski, I. M. Besieris, and A. M. Shaarawi, "Localized wave representations of acoustic and electromagnetic radiation," Proceedings of the IEEE, Vol. 79, No. 10, pp. 1371–1377, October, 1991

LEGENDS

**Fig. 1.** Envelope detected Zeroth-order X Wave in a 50 mm diameter rigid acoustic waveguide. The waves shown has an Axicon angle of $4^o$ and $a_0 = 0.05$ mm. (a) and (c) Band-limited version with a Blackman window function centered at 3.5MHz with about 81% of fractional -6dB bandwidth. (b) and (d) are a broadband version. The images in the top row are in a linear scale and those in the bottom row have a log scale to show the sidelobes.

**Fig. 2.** The same as those in Fig. 1 except that the images are zoomed horizontally around the center.

**Fig. 3.** Transverse ((1) and (3)) and axial sidelobe ((2) and (4)) plots of the images in Fig. 1. ((1) and (2)) and Fig. 2. ((3) and (4)), respectively. Solid and dotted lines are for band-limited and broadband cases.

FIGURES

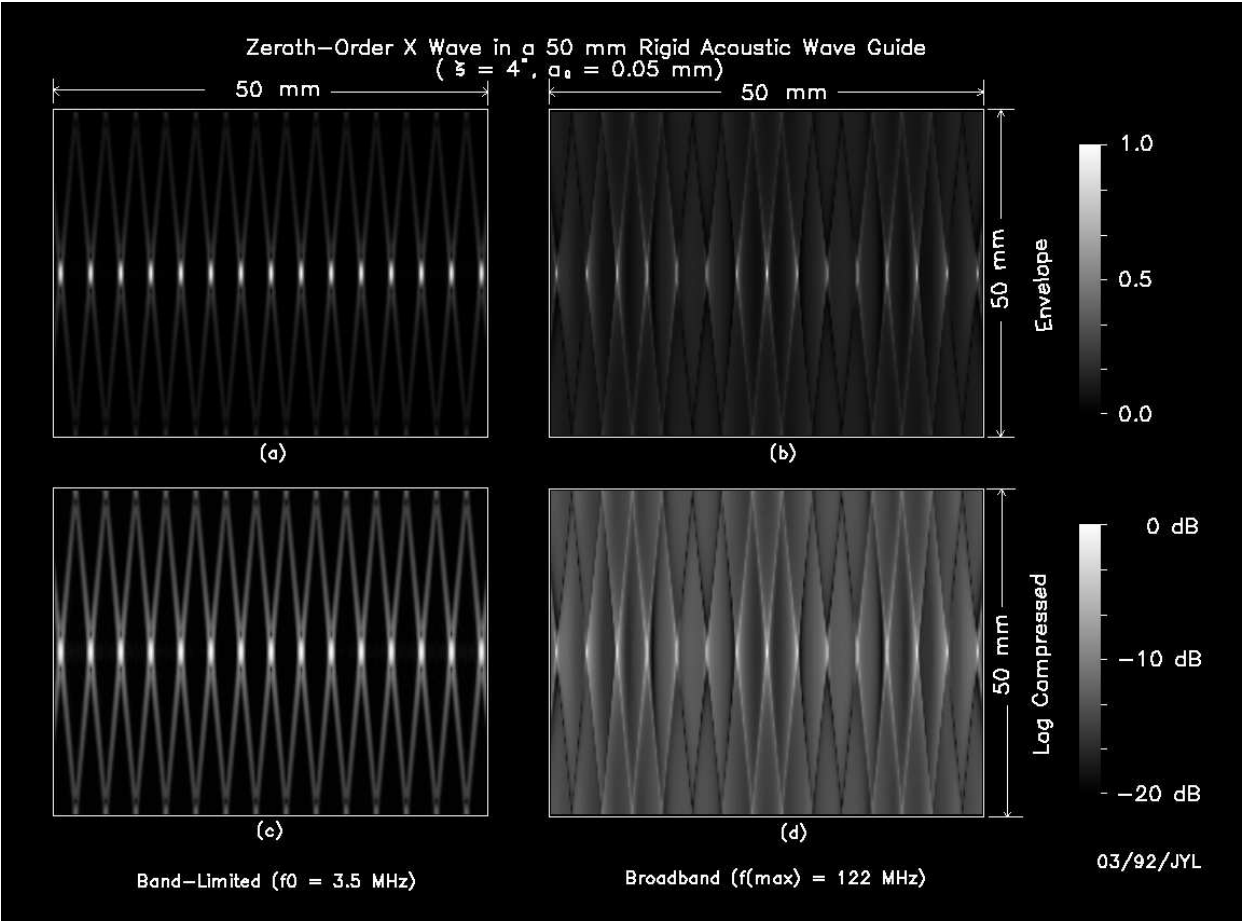

Fig. 1.

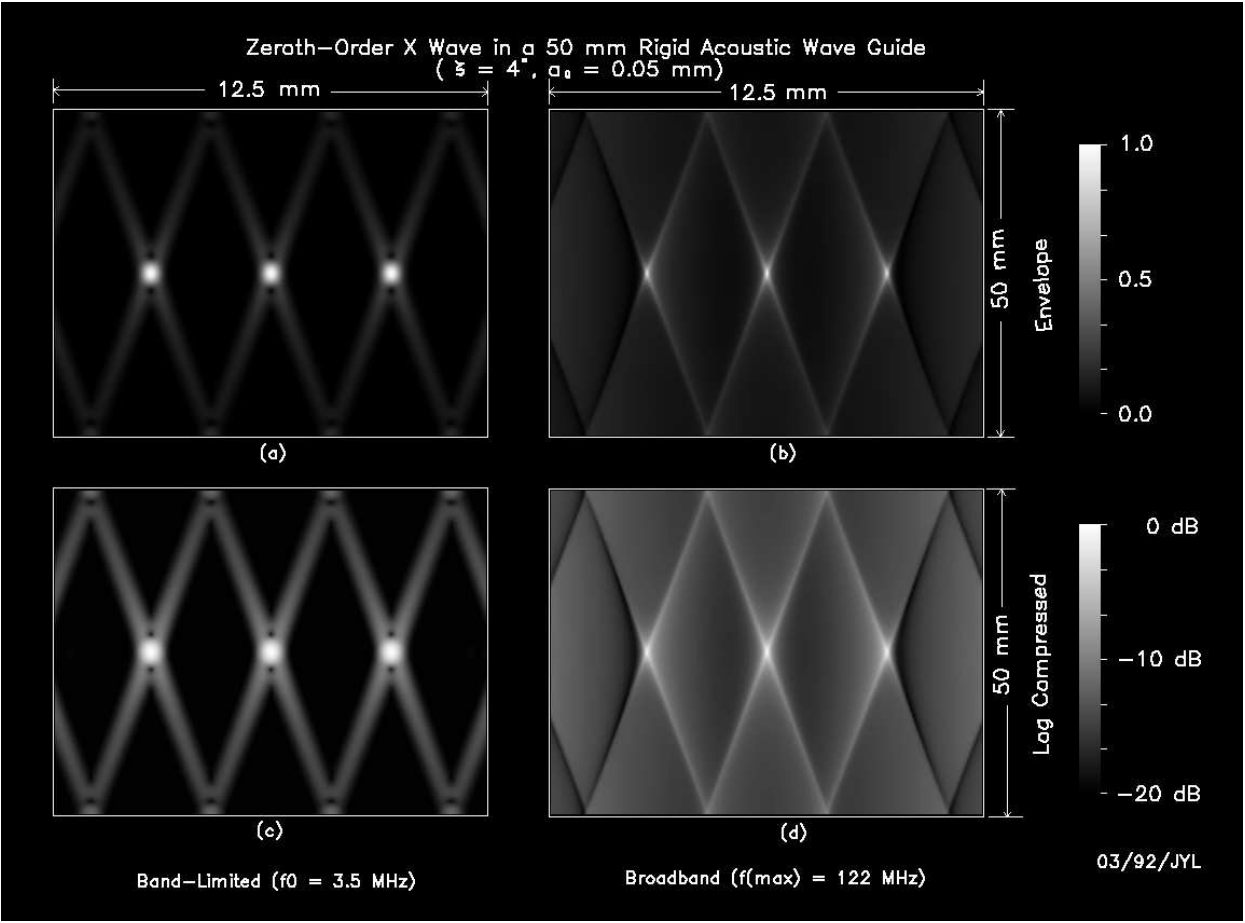

Fig. 2.

# X WAVE IN A 50 MM WAVE GUID

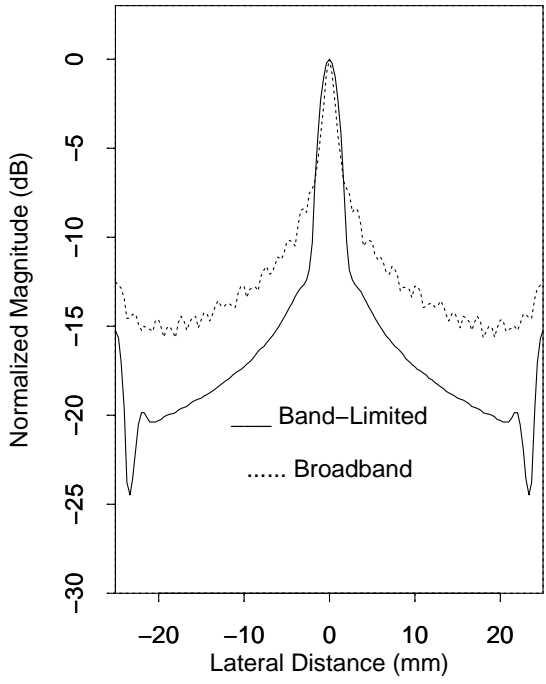

(1)

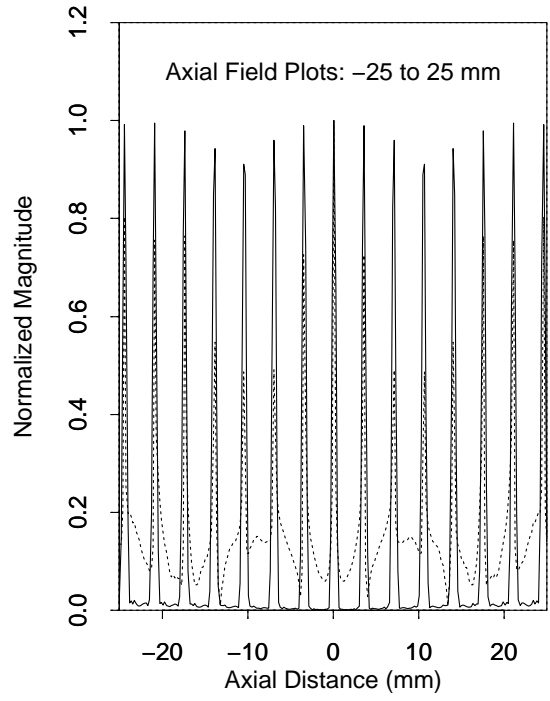

(2)

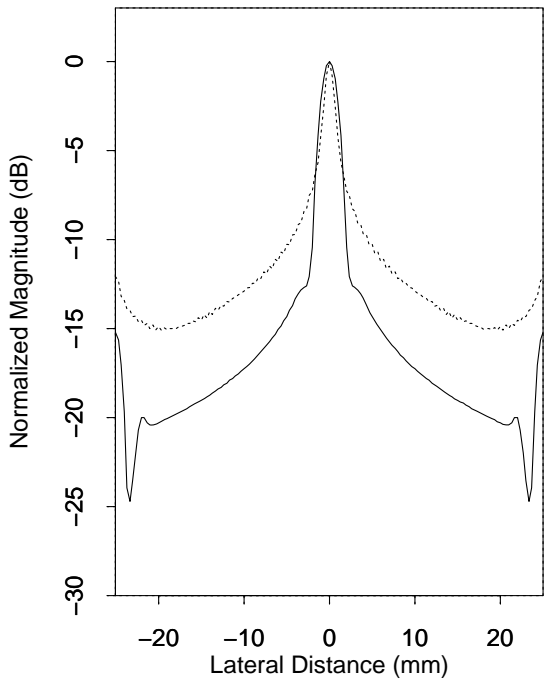

(3)

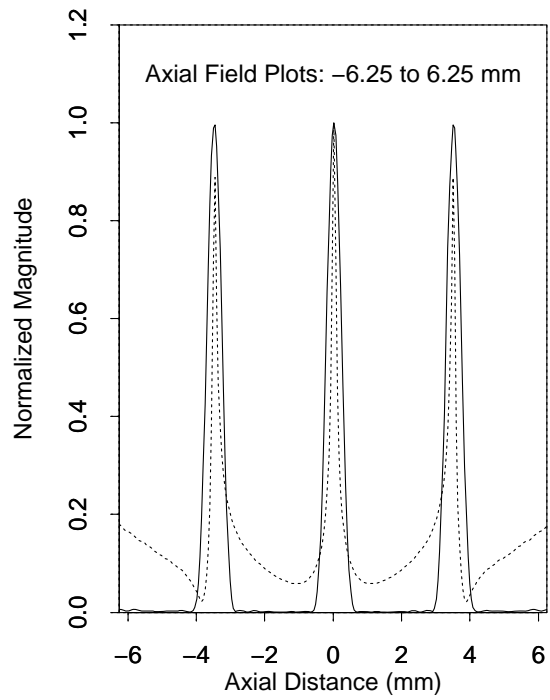

(4)

Fig. 3.